\documentclass[twocolumn,nofootinbib]{revtex4-1}
\usepackage{latexsym,epsfig,amssymb, amsmath,nicefrac}
\usepackage{epsfig,graphicx}
\usepackage[usenames, dvipsnames]{color}
 
              \newcommand{\rf}[1]{(\ref{#1})}

\def\bfone{\relax{\rm 1\kern-.35em 1}}

\newcommand{\be}{\begin{equation}}
\newcommand{\ee}{\end{equation}}
\newcommand{\ben}{\begin{displaymath}}
\newcommand{\een}{\end{displaymath}}
\newcommand{\bea}{\begin{eqnarray}}
\newcommand{\eea}{\end{eqnarray}}

\newcommand{\bean}{\begin{eqnarray*}}
\newcommand{\eean}{\end{eqnarray*}}

\newcommand{\vp}{\varphi}

\def\K{K{\"a}hler}

\makeatletter \@addtoreset{equation}{section} \makeatother

\begin{document}
\title{\Large{On inflation, cosmological constant, and SUSY breaking}}

\author{Andrei Linde}

\affiliation{Department of Physics and SITP, Stanford University, \\ 
Stanford, California 94305 USA}

\begin{abstract}
We consider a broad class of inflationary models of two unconstrained chiral superfields, the stabilizer $S$ and the inflaton  $\Phi$, which can describe inflationary models with nearly arbitrary potentials. These models include, in particular, the recently introduced theories of cosmological attractors, which provide an excellent fit to the latest Planck data. We show that by adding to the superpotential of the fields $S$ and  $\Phi$ a small  term depending on a nilpotent chiral superfield $P$ one can break SUSY and introduce a small cosmological constant without affecting main predictions of the original inflationary scenario. 
\end{abstract}

\maketitle

\smallskip

\section{Introduction}

There was a significant progress in designing cosmological models in supergravity using absolutely minimal number of ingredients. 
For example, the very first version of chaotic inflation in  supergravity, GL model  \cite{Goncharov:1983mw,Linde:2014hfa},  describes a single chiral superfield with a plateau potential in the inflaton direction. Along with  the Starobinsky model       \cite{Starobinsky:1980te},  the Higgs inflation \cite{Salopek:1988qh,Bezrukov:2007ep}, and a broad class of the cosmological attractors to be discussed below, GL model provides one of the best matches to the Planck data \cite{Planck:2015xua}. 

During the last two decades, attention switched to models of two chiral superfields, the stabilizer $S$ and the inflaton  $\Phi$ \cite{Kawasaki:2000yn,Kallosh:2010ug}. Such models are very simple and flexible and can have nearly arbitrary inflationary potentials \cite{Kallosh:2010ug}. The basic idea is to consider models where the \K\ potential of the field $\Phi$ has a flat direction, such as $\phi =\Phi-\bar\Phi$, or $\phi = \Phi+\bar\Phi$, corresponding to the inflaton field $\phi$. Then by choosing a superpotential $W = S f(\Phi)$ one finds a broad class of models which can have nearly arbitrary inflationary potentials $V =| f(\phi)|^{2}$ \cite{Kallosh:2010ug}. 

Such models include, in particular, the recently introduced theories of $\alpha$-attractors, which have an advantage of successfully matching the existing Planck data nearly independently of the initial choice of the inflationary potential, see  \cite{Kallosh:2013hoa} and references therein. These models have a beautiful geometric origin \cite{Kallosh:2015zsa,Carrasco:2015uma} protecting asymptotic flatness of the potential \cite{Kallosh:2016gqp},  and they simplify the solution of the initial conditions problem for inflation \cite{Carrasco:2015pla}.

All of the models described above have a common feature: After inflation in these models, supersymmetry is unbroken and vacuum energy vanishes. This could seem to be an excellent approximation: most of the phenomenological models presume that SUSY breaking and the vacuum energy are extremely small, $m_{3/2} \ll 10^{-10} $,  and $\Lambda \sim 10^{{-120}}$ in the units $M_{p} = 1$. However, there is a no-go theorem saying that one cannot obtain models with a small SUSY breaking and a small positive cosmological constant by a minor change of parameters of the models  with unbroken SUSY in Minkowski vacuum \cite{Kallosh:2014oja}. 

\newpage

The most elegant way to introduce a small SUSY breaking and a small positive cosmological constant in these models is based on the concept of orthogonal nilpotent superfields, which essentially makes $S = {\rm Im} \Phi = 0$. The resulting models \cite{Ferrara:2015tyn} are almost as general as the original models    \cite{Kallosh:2010ug,Kallosh:2013hoa}, and they are more economical since they remove the fields $S$ and ${\rm Im} \Phi$ from consideration, which makes all calculations nearly trivial. 

However,  this powerful method does not apply to some models, such as the single-field GL model  \cite{Goncharov:1983mw,Linde:2014hfa}. Similarly, the supergravity implementation of the Higgs inflation  \cite{Ferrara:2010yw} does not allow to use the constraint ${\rm Im} \Phi = 0$, and the recent versions of sneutrino inflation  \cite{Nakayama:2016gvg,Kallosh:2016sej}, require  the field $S$ to remain a dynamical, unconstrained chiral superfield.
In such situations one may consider more traditional methods of symmetry breaking, e.g. by adding to the theory a Polonyi-like field $P$. In the past, such fields have been the source of the cosmological light moduli problem, which plagued supergravity cosmology for more than 30 years \cite{Coughlan:1983ci}. Fortunately, this problem does not appear if this field is nilpotent or strongly stabilized. 

In particular, it was found that by adding a strongly stabilized or nilpotent chiral Polonyi-type superfield $P$ to the GL model and its  generalizations,  one can introduce a small SUSY breaking and a cosmological constant without making any noticeable changes to its   inflationary predictions \cite{Linde:2014hfa,Kallosh:2015lwa,Roest:2015qya}. Similar approach can work in the theories with the inflaton in the massive vector multiplet \cite{Farakos:2013cqa,Ferrara:2013rsa,Ferrara:2014cca,Aldabergenov:2016dcu}.

In this paper I will show that the same method can be used for the broad class of the models \cite{Kallosh:2010ug,Kallosh:2013hoa}.  By adding to the superpotential of the fields $S$ and  $\Phi$ a small  term depending on a nilpotent chiral superfield $P$ one can break SUSY and introduce a small cosmological constant without affecting main predictions of the original inflationary scenario. 
Instead of trying to develop the most general class of models of this type, I will describe two most representative models which will clearly demonstrate how this method works.

\section{The simplest example: Chaotic inflation with a quadratic potential, SUSY breaking and a cosmological constant}

The simplest version of   inflation in supergravity developed in  \cite{Kawasaki:2000yn} has the \K\ potential and superpotential which can be represented as follows:
\be
K=-{(\Phi- \bar\Phi)^2\over 2} + S\bar S  \ ,\qquad  W=  m\ S\,  \Phi   \ .
\label{chaotyan}
\ee
We will represent the scalar fields $S$ and $\Phi$ in terms of their canonically normalized components $S =(s+ia)/\sqrt 2$ and $\Phi = (\phi+i\chi)/\sqrt 2$. During inflation in this model, $S = \chi = 0$, and the field $\phi$ plays the role of the inflaton field, with the  potential ${m^{2}\over 2}\phi^{2}$. In the minimum of the potential one has $\phi = 0$, $V = 0$, and the gravitino mass, representing the strength of SUSY breaking is $m_{3/2} = 0$.

The modification of this model that we propose is
\be
K=-{(\Phi- \bar\Phi)^2\over 2} + S\bar S + P\bar P  \ ,
\ee
 \be
 W=  m\ S\,  \Phi  + (\sqrt 3\, P + 1 )\, m_{3/2}  \ .
\label{twogolda}\ee

Investigation of the potential $V(S,\Phi)$ in this theory shows that, just as in the general class of theories studied in \cite{Kawasaki:2000yn,Kallosh:2010ug} with the superpotential $W=  m\ S\, f(\Phi )$, the imaginary parts of the fields $S$ and $\Phi$ vanish during inflation. Moreover, both imaginary and real parts of these fields vanish after inflation. However, the real part of the field $S$ no longer vanishes during inflation.

Indeed, the potential for non-vanishing canonically normalized scalar components $s$ and $\phi$ for $a=  \chi=0$ is 
\bea
V &=&  \frac{m^2}{2} e^{\frac{s^2}{2}} \Bigl(\phi^2 +s^{2}\Bigl(1- {2- s^2\over 4} \phi^2 \Bigr) \nonumber\\
&+ &  {m_{{3/2}}\over  m} s \Bigr((s^{2} - 4)\phi+{m_{{3/2}}\over m} s\Bigr)\Bigr) \ .
\label{potgold5a}\eea
The main effect  of the term $(\sqrt 3\, P + 1 )\, m_{3/2}$ in the superpotential on the dynamics of inflation is the shift of the position of the minimum of the potential with respect to the field $s$. Throughout the paper we will assume that  ${m_{{3/2}}\over m} \ll 1$. For $m_{{3/2}}=0$  the minimum is at $s = 0$, but for $m_{{3/2}}\not =0$ it is slightly displaced by
\be
s =   2 \phi {m_{{3/2}}\over m} \ ,
\ee
up to the terms higher order in the small parameter ${m_{{3/2}}\over m}$.  The inflaton potential along the shifted inflationary trajectory becomes 
\be
V =  \frac{\phi^2}{2} (m^{2} -4m_{{3/2}}^{2}) \ . 
\ee
This trivial modification does not affect the inflationary dynamics for $m_{{3/2}} \ll m$, but now supersymmetry becomes broken in the minimum of the potential.

In order to describe a small cosmological constant, one can add a tiny correction to the superpotential,
 \be
 W=  m\ S\, f(\Phi) + (\sqrt {3+\delta}\, P + 1 )\, m_{3/2}  \ .
\label{twogoldb}\ee
This correction leads to the cosmological constant, 
\be\label{CC}
V_{0} = \delta\, m_{3/2}^{2} \ ,
\ee
up to higher order corrections in $m_{3/2}$. For $V_{0} \sim 10^{{-120}}$, the corresponding modification of the potential results in a negligible modification of the inflationary scenario.

For a more general scenario with
\be
 W=   S\,  f(\Phi)  + (\sqrt {3+\delta}\, P + 1 )\, m_{3/2}  \ ,
\label{gen}\ee
in the first approximation in the small parameter ${m_{{3/2}}\over m}$ the field $s$ has a minimum at
\be
s =  2\sqrt 2 m_{3/2} {f({\phi\over \sqrt 2})\over (f'({\phi\over \sqrt 2}))^{2}}
\ee
and the potential is
\be
V = f^{2}\bigl({\phi\over \sqrt 2}\bigr)\left(1-{4m_{{3/2}}^{2}\over (f'({\phi\over \sqrt 2}))^{2}}\right) \ .
\ee

\section{$\alpha$-attractors }
Now we will consider the simplest  $\alpha$-attractor model with the \K\ potential vanishing along the inflaton direction $\Phi= \bar\Phi$ \cite{Carrasco:2015uma}:
\be\label{KZ}
K= -{3  \alpha\over 2}  \log  \left[ {( 1- \Phi \overline{\Phi} )^2  \over  (1-\Phi^2) (1- \overline{\Phi}^2)}  \right] + S \overline{S}  ,  
\ee
and the same superpotential as before,
\be
 W=  m\ S\,  \Phi  \ .
\label{t1}
\ee
In this theory, the inflaton potential is also quadratic with respect to the field ${\rm Re}\, \Phi$, but this field is not canonical. After the transition to the canonically normalized field $\varphi$, such that 
\be \label{tr}
 {\rm Re}\, \Phi = \tanh { \vp\over \sqrt {6\alpha}}, 
\ee
the inflaton potential becomes 
\be\label{tmod}
V =  m^2 \tanh^2{ \vp\over \sqrt {6\alpha}}.
\ee
This is a plateau potential representing the simplest example of what we called T-model \cite{Kallosh:2013hoa}. Once again, at the minimum of the potential all fields vanish, the potential vanishes, and supersymmetry is unbroken. 

Now we will generalize this model in the same way as we did in the previous section:
\be\label{KMATTER}
K= -{3  \alpha\over 2}  \log  \left[ {( 1- \Phi \overline{\Phi} )^2  \over  (1-\Phi^2) (1- \overline{\Phi}^2)}  \right] + S \overline{S} + P \overline{P}  ,  
\ee
and
\be
 W=  m\ S\,  \Phi  + (\sqrt {3+\delta}\, P + 1 )\, m_{3/2}  \ .
\label{twogolda1}
\ee

At small $\Phi$, investigation of this model is reduced to the one performed in the previous section. This means that our results for the position of the minimum of the potential at $\Phi = 0$, $S = 0$,  the gravitino mass, and the expression for the cosmological constant \rf{CC} remain intact. Investigation of inflation is unaffected by the tiny parameter $\delta$. Therefore in what follows we will concentrate on the simplest  case $\delta = 0$. 

As before, one can show that the   term $(\sqrt {3}\, P + 1 )\, m_{3/2}$ does not induce imaginary part of the fields $\Phi$ and $S$. Therefore one can replace $S$ by its canonically normalized real component,  $S =s/\sqrt 2$, and $\Phi$ by the canonically normalized inflaton field $\varphi$ \rf{tr}.

In terms of $\varphi$ and $s$, the potential is given by
\bea
V &=&  e^{\frac{s^2}{2}}  m^2 \Bigl[\tanh ^2 \frac{\varphi }{\sqrt{6\alpha} }  +{s^2\over 12 \alpha} \Bigl(2  \tanh ^4\frac{\varphi }{\sqrt{6\alpha}}+2  \nonumber\\ &+&\bigl(3 \alpha  s^2-2 (3 \alpha +2)\bigr) \tanh ^2 \frac{\varphi }{\sqrt{6\alpha} } \Bigr)\nonumber\\  &+&\Bigl(\frac{s\, m_{3/2}}{\sqrt 2 m}\Bigr)^2 -\frac{s\, m_{3/2}}{\sqrt 2 m}   (4-s^2)  \tanh  \frac{\varphi }{\sqrt{6\alpha}} \Bigr] .
\eea
This expression looks quite complicated, but only the last of its terms, suppressed by the small ratio $\frac{m_{3/2}}{\sqrt 2 m}$, shifts the field $s$ from 0. If this term is small, the field $s$ is small, and the potential returns to  its simple original expression \rf{tmod}.

It is instructive to evaluate  the magnitude of the  shift of the field $s$ away from its original minimum at $s = 0$ induced by the small  gravitino mass. For simplicity, we will do it  in the limit $\varphi \gg \sqrt {6\alpha}$, representing the main part of the inflationary plateau in the theory of $\alpha$ attractors. In this limit $\tanh{ \vp\over \sqrt {6\alpha}} \to 1$, and the potential becomes
\be
V =  e^{s^{2}\over 2} m^{2}\Bigl(1 -{s^{2} \over 2} + {s^{4}\over 4} -
(4 -s^{2} ){s\, m_{3/2} \over  m \sqrt 2} + \Bigl({s\, m_{3/2} \over  m \sqrt 2}\Bigr)^{2}\Bigr)
\ee
For $m_{3/2} \ll  m$, the minimum of this potential with respect to $s$ is at 
\be
s = 2^{{5/6}}\, \Bigl({m_{3/2} \over  m}\Bigr)^{1/3} \ .
\ee
For $\varphi \gg \sqrt {6\alpha}$, the shift of the field $s$ from its original state at $s = 0$ is even smaller, and at the global minimum of the potential one has $s = \varphi = 0$. Therefore for $m_{3/2} \ll  m$ the inflaton potential with a good accuracy is represented by its original expression \rf{tmod}. The main modification of this potential during inflation is a slight decrease of its height,
\be
V(\varphi \to \infty) \approx m^{2}\Bigl(1- 6 \Bigl({m_{3/2} \over  \sqrt 2\, m}\Bigr)^{4/3}\Bigr) \ .
\ee
To give a particular numerical example, one may consider the realistic case with $m \sim 10^{{-5}}$ and $m_{3/2} \sim 10^{{-14}}$ in Planck mass units, i.e. $m_{3/2} \sim 20 $ TeV. This leads to the value of the field $s$ along the inflationary plateau $s \sim 2\times 10^{{-3}}$, and the height of the potential decreasing by $10^{-12}$ of its original value. These tiny changes do not affect inflationary observational predictions of these models.

\section{Discussion}
In this paper, we assumed that the Polonyi-type field P is nilpotent, which means that its scalar component and its scalar perturbations vanish. This significantly simplifies the investigation, and immediately solves the cosmological Polonyi field problem, which plagued supergravity cosmology for 3 decades \cite{Coughlan:1983ci}. We expect that similar results should remain valid if Polonyi fields are unconstrained but strongly stabilized \cite{Dudas:2012wi,Abe:2014opa}. 
However,  models with  nilpotent  fields may be advantageous because of their possible string theory interpretation  \cite{Ferrara:2014kva}.

In the investigation above we did not use any stabilization of the field $S$ near $S = 0$. This is the simplest case, and it is encouraging that even in this case inflation is not much affected by SUSY breaking. Note that in this simple scenario, the mass of the field $S$ during inflation is very small, so that in addition to the inflaton perturbations, the perturbations of the field $S$ are also generated. However, one can typically ignore these perturbations. They do not lead to observable adiabatic or isocurvature perturbations  \cite{Kallosh:2013daa}, unless one considers special cases where the decay rate of the $S$ field is extremely strongly suppressed \cite{Linde:1984ti,curva,Demozzi:2010aj}, or the field $S$ strongly interacts with the inflaton field, which may modulate duration of inflation and the process of reheating \cite{Dvali:2003em}. 
If one stabilizes the field $S$ by adding terms $\sim S^{4}$ to the \K\ potential,  the perturbations of the field $S$ disappear, and the modifications of the inflationary trajectory induced by the SUSY breaking terms become even smaller than in the simplest scenario without stabilization.   

Thus one can use the full functional freedom and flexibility of construction of inflationary models of this class without SUSY breaking and uplifting \cite{Kallosh:2010ug,Kallosh:2013hoa}, and delegate the role of uplifting and SUSY breaking to the nilpotent (or strongly stabilized) Polonyi-type field. The models of this class are conceptually simple and can successfully describe inflation, cosmological constant, and SUSY breaking.

I am grateful to  S. Ferrara, R. Kallosh, D. Roest, T. Wrase, and Y.~Yamada for  enlightening  discussions.   This work  is supported by the SITP,  and by the NSF Grant PHY-1316699.

\end{document}